\documentstyle[prl,aps,epsf,floats,twocolumn]{revtex}
\begin{document} 
\newcommand{\be}{\begin{equation}}
\newcommand{\ee}{\end{equation}}
\newcommand{\TS}[1]{{\tt <T>}{\bf #1}{\tt </T>}}
\newcommand{\AS}[1]{{\tt <A>}{\bf #1}{\tt </A>}}

\title{Testing for nonlinearity in unevenly sampled time series}

\author{Andreas Schmitz and Thomas Schreiber\\
      Physics Department, University of Wuppertal,\\
      D--42097 Wuppertal, Germany}
\maketitle
\begin{abstract} 
  We generalize the method of surrogate data of testing for
  nonlinearity in time series to the case that the data are sampled
  with uneven time intervals. The null hypothesis will be that the
  data have been generated by a linear stochastic process, possibly
  rescaled, and sampled at times chosen independently from the
  generating process. The surrogate data are generated with their
  linear properties specified by the Lomb periodogram. The inversion
  problem is solved by combinatorial optimization.
\end{abstract}

\vspace{0.5cm}

The vast majority of methods of time series analysis deals with data
measured at times which are an integer multiple of the fixed sampling
interval $\Delta$. Unevenly sampled time series are often excluded
although they are quite common in cases where measurements are
restricted by practical conditions. For example, most astronomical
observations cannot be made in the day time and must often be
interrupted in the night due to cloudy weather; data from the stock
exchange has gaps during the weekends and holidays, etc. The reason
for excluding these is mainly technical: most methods cannot be easily
generalized to this case. This is particularly true for nonlinear
methods of time series analysis~\cite{book}.

For the nonlinear approach to a given time series, we first have to
ask for signatures of nonlinearity in the generating process. In this
paper for the first time we present a method to test for nonlinearity
in unevenly sampled time series. We will extend the common concept of
surrogate data~\cite{surro}, where randomized data sets are used to
obtain a Monte Carlo approximation to the probability distribution of
a suitable test statistics.

For this approach, we have to be able to generate sequences that are
random except for their linear correlations. Any additional structure
can lead to spurious positive results. The null hypothesis in this
paper is that the data have been generated by a linear stochastic
process that is measured instantaneously and possibly rescaled. This
excludes correlations or even a deterministic relationship between the
sampling intervals and the observable. This method can however be
generalized to that case.

Time series taken at varying time intervals are very difficult to
handle and only few successful efforts have been made in this area.
In certain situations, one can interpolate the data to equally spaced
sampling times. However, in a test for nonlinearity, one could then
not distinguish between genuine structure and nonlinearity introduced
spuriously by the interpolation process.

If we want to generate surrogate data sets with the same linear
properties as the original data, the autocorrelations have to be
conserved. But even such simple things like autocorrelations are hard
to maintain for unevenly sampled data. Any time interval can occur
between successive points and it is possible to combine them to nearly
arbitrary lags.  One idea is to calculate autocorrelations by binning
all possible intervals to the desired, discrete lags, a process that
involves some nonlinearity. Using these autocorrelations for
generating surrogates can lead to the spurious rejection of purely
linear time series.

Besides the generation of surrogates, we have to be able to measure
the degree of nonlinearity in the data. In contrast to the process of
generating surrogates, interpolations are permitted here as part of
the specification of a test statistic. A badly designed test statistic
could in the worst case lower the discrimination power of the test,
while still keeping it formally correct. In this paper we use a very
simple test statistic that measures nonlinearity through deviations
from time--reversibility. The main emphasis is laid on the generation
algorithm for the surrogates.

Standard surrogate methods make use of the Fourier transformation to
conserve the autocorrelations of the original data. The method of {\em
  amplitude adjusted Fourier transformation} (AAFT,~\cite{surro})
rescales the original time series to a Gaussian distribution first.
Then, the Fourier phases are randomized and the Fourier transformation
is inverted. Finally, a rescaling to the original distribution is
performed. A refined method has been suggested in Ref.~\cite{we} where
an iteration scheme is used to simultaneously conserve the spectrum
and the distribution. It consists of alternating Fourier
transformation and rescaling steps. Both methods cannot directly be
applied to unevenly sampled data, because they utilize the Fourier
transformation and its inverse.

In Ref.~\cite{anneal_surr}, a general approach to the constrained
randomization of time series is described. There, a desired property
of the surrogate data is expressed through a cost function. The
minimum of this cost function is reached if the surrogate fulfills the
given property. In this paper, we will make use of this method with a
cost function that can also be defined for unevenly sampled data. This
will enable us to impose the desired linear correlation structure on
the surrogate time series.

Let $\{y_n\}$ be a time series sampled at times $\{t_n\}$ that need
not be equally spaced. The Power spectrum can then be estimated by the
Lomb periodogram~\cite{lomb}. This spectral estimator is discussed
e.g. in Ref.~\cite{NR_lomb}. Here we give the final formula:
\begin{eqnarray}
  \label{LS}
  P(\omega)=\frac{1}{2\sigma^2}\left\{
    \frac{{[\sum_n(y_n-\bar{y})\sin\omega (t_n-\tau)]}^2}
    {\sum_n\sin^2\omega(t_n-\tau)}+ \right. \nonumber \\
    \left. \frac{{[\sum_n(y_n-\bar{y})\cos\omega (t_n-\tau)]}^2}
    {\sum_n\cos^2\omega(t_n-\tau)}
  \right\}
\end{eqnarray}
where $\tau$ is defined by
\begin{equation}
  \label{tau}
  \tan(2\omega\tau)=\frac{\sum_n\sin 2\omega t_n}{\sum_n\cos 2\omega t_n}
\end{equation}
and $\bar{y},\sigma^2$ are the mean and the variance of the data,
respectively. The result can be derived by fitting a least squares
model $y_n=a\cos\omega t_n + b\sin\omega t_n$ to the data for each
given frequency $\omega$. Therefore Lomb periodograms are often
referred to as {\em least squares periodograms}.

For time series sampled at constant time intervals,
$\Delta=t_n-t_{n-1}$ for all $n$, the Lomb periodogram $P(\frac{2\pi
  n}{N\Delta})$ yields the standard squared Fourier transformation.
Except for this particular case, there is no inverse transformation
for the Lomb periodogram, which makes it impossible to use the
standard surrogate data algorithms mentioned above. Therefore, we
follow the general approach of Ref.~\cite{anneal_surr}, where
constraints on the surrogate data are implemented by a cost function
$E(\{y_n\})$ which has a global minimum when the constraint is
fulfilled. As in Ref.~\cite{anneal_surr} this cost function will be
minimized by simulated annealing~\cite{kirk,kirk2}. Starting with a
random permutation of the original time series, the surrogate is
modified by exchanging two points chosen at random.  The modification
will be accepted if it yields a lower value for the cost function or
else with a probability $p=\exp(-\Delta E/T)$.  The ``system
temperature'' $T$ will be lowered slowly to let the system settle down
to a minimum. This idea goes back to Metropolis et al.~\cite{metro}.
See for example~\cite{anneal} for details.

In our case, we use the Lomb periodogram of the data as a constraint
for the surrogates. It can be expressed as a cost function for example
by:
\begin{equation}
  \label{cost}
  E=\left[
    \sum_{k=1}^{N_f} |P(k\omega_0)-P_{data}(k\omega_0)|^q
  \right]^{1/q}
\, .
\end{equation}
We use $P$ at $N_f$ equally spaced frequencies $k\omega_0$, other
choices are possible. The parameter $q$ specifies the distance
measure between the two periodograms. For $q=2$, the $L^2$--distance is
used. For the following examples we use $q=1$. Higher ``penalties''
for large differences in single frequencies could be given by raising
$q$ above 1.  Alternatively, one could use the differences between the
square roots or logarithms of $P$, which puts less stress on the peaks
in the power spectrum than (\ref{cost}). Another freedom lies in the
choice of the minimum frequency $\omega_0$ and the number of
frequencies $N_f$ and one may have to consider different values for
each individual application.

Recall that we are interchanging only the values of the time series
$y_n$, while fixing the times $t_n$ where measurements are made. This
excludes data like inter--beat intervals of electrocardiograms where
we have the additional condition $t_{n+1}=t_n+y_n$.

For the calculation of surrogates, simulated annealing is performed
until $E$ has fallen below a given value $E_f$, the desired accuracy
of the Lomb periodogram. In Monte Carlo simulations, new
configurations are usually generated from ``older'' configurations. In
order to avoid correlations between the different surrogates, we
prefer to start with a completely random permutation of the original
time series for each surrogate. The starting temperature can roughly
be determined by calculating the cost function for some randomly
shuffled data sets and choosing $T_0$ as a ``typical'' difference in
cost between them. Once an adequate starting temperature is known, it
can be used for further surrogates.

Two improvements that accelerate the annealing algorithm have been
made. The first is to choose the two points that are candidates for an
exchange with a probability that depends on their difference in
magnitude. Exchanging two points with a big difference in their rank
(eg. the smallest and the largest value) yields a larger change of the
cost function than exchanging two points that do not differ much in
their ranks. For good performance, it is desired to keep approximately
the same acceptance rate through all temperatures. This can be
achieved by choosing pairs of points $\{y_i,y_j\}$ with $i\neq j$ and
probability $p_{ij}(d,\mu)$ where $d=|{\rm rank}(y_i)-{\rm
  rank}(y_j)|-1$. The probability $p$ is chosen to have a maximum for
$d=0$ and to decrease for higher $d$. The parameter $\mu$
characterizes the ``width'' of the distribution $p$ and should be
varied proportional to $N/T$. The exact shape of $p$ does not seem to
be of much importance. For example, we were not able to observe a
significant difference in performance between exponential and Gaussian
distributions. But in all cases we considered, a non--uniform $p_{ij}$
substantially accelerated the annealing process, so that higher
accuracy is reachable with the same computational effort.

Calculating $E$ is very time consuming for long time series and many
frequencies. With a typical value of $N_f \propto N$ we have an
algorithm of order $N^2$ for each annealing step. Additionally, the
number of annealing steps grows at least linearly with $N$. For our
applications, it is not necessary to recalculate all sums in
(\ref{LS}) for every exchange, because we only change the values $y_n$
while fixing the times $t_n$ and frequencies $k\omega_0$. In
(\ref{LS}), $\tau$ and the two denominators do not depend on $y_n$ and
can be stored in arrays for every frequency $k\omega_0$ at the
beginning of the annealing process. The sums in the numerator do not
change much either and only two terms have to be recalculated. This
reduces the recalculation of the Lomb periodogram to order $N$.

But even with the described modifications to the algorithm the actual
annealing is quite computer time intensive. Generating one surrogate
for a time series with $N=2000$ points while fixing $N_f=1000$
frequencies takes about 5 hours CPU--time on a DEC alpha workstation
at 166~MHz.

So far, we have described how to produce randomized versions of
unevenly sampled time series with given linear correlations, which is
the main point of this paper. Let us now demonstrate, how such
surrogate sequences can be used in tests for nonlinearity. For this
purpose, we have to be able to measure the degree of nonlinearity in a
time series. Many statistics that have proven useful for evenly
sampled time series (see e.g.~\cite{dispow}) cannot easily be
generalized to unevenly spaced data. This generalization, in general,
is a topic of future research.

Here, as a first simple statistic, we choose a measure for
time--reversibility, which is a good indicator for nonlinearity. It is
however not very enlightening about what source of nonlinearity there
might be. For the data sorted in time order,
\begin{equation}
\label{trev}
\gamma=\frac{1}{(\sigma^2)^{\frac{3}{2}}(N-1)} \sum_{n=2}^{N}
\left( \frac{y_n - y_{n-1}}{t_n - t_{n-1}} \right) ^3
\end{equation}
is calculated, which is just the mean of the slopes, taken to the
third power. For a time series generated by a linear process, and for
the surrogates, we expect $\gamma \approx 0$. In contrast, time series
with nonlinearities can be asymmetrical in time and may yield values
of $\gamma \neq 0$. To pay regard to deviations in both directions
($\gamma>0$ and $\gamma<0$), a {\em two sided} test~\cite{constr} has
to be performed.

\begin{figure}
\centerline{
\mbox{\hspace{1.9cm}\epsffile{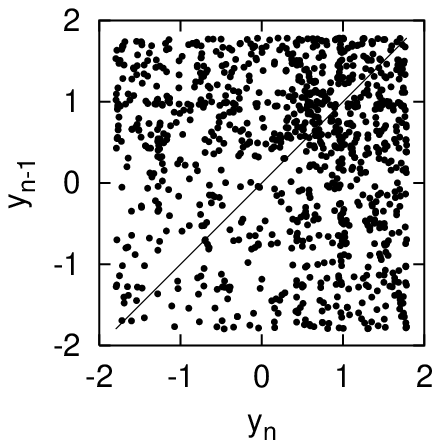}\hspace{-2.4cm}\epsffile{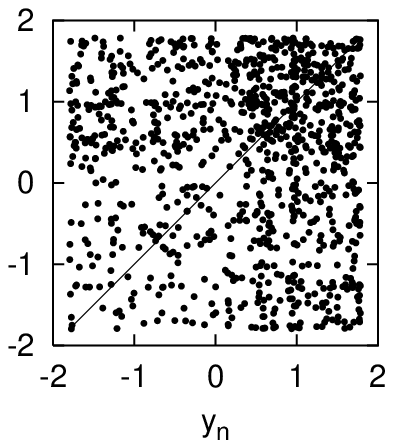}}}
\caption[]{\label{fig:phase}
  Delay--plot of an unevenly sampled H\'enon map and one surrogate.
  The test finds a significant difference in time--asymmetry.}
\end{figure}

To test the functionality of the surrogate test, we use 10000 points
of the H\'enon map as a first example. From these, we pick $N=1000$
points with their time indices chosen randomly. To generate
surrogates, we calculate the Lomb periodogram for $N_f=500$
frequencies in the interval $\nu \in [0, 0.5]$. A delay--plot of the
data and one surrogate is shown in Fig.~\ref{fig:phase}. A little
trace of the H\'enon attractor can be found in the left figure that is
built by pairs of values with time delay one. For the original time
series we get $\gamma=-0.58$, while 19 surrogates gave values $-0.30 <
\gamma < 0.18$. This corresponds to a $90 \%$ level of significance
for the time series not being time reversible and hence nonlinear in
the sense of the null hypothesis.

In order to verify the new test method we also applied the test to
linear time series based on an AR(1)--process $x_{n+1}=0.95 x_n +
\eta_n$ which is however invertibly rescaled by $y_n=x_n \sqrt{x_n}$.
Again we picked 1000 points randomly from the original 10000. A test
performed with this data was unable to reject the null hypothesis, as
expected since the null was true.

For a quantification of the power and the size~\cite{constr} of the
new method, many independent tests would be necessary. Such an
extremely computer time--intensive study is beyond the scope of the
present work.

\begin{figure}
\hspace{-0.6cm}\epsffile{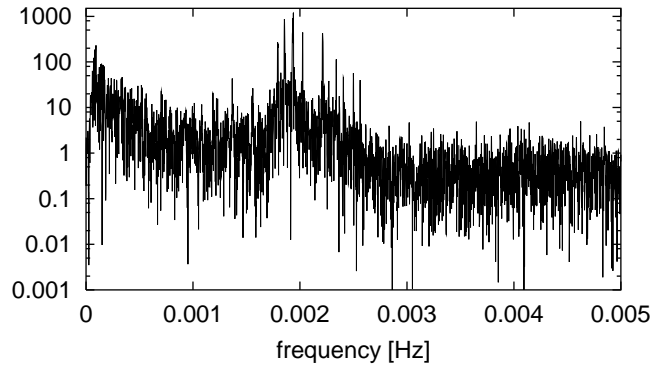}
\caption[]{\label{fig:elomb}
  Lomb periodogram of data set E. See text for details. }
\end{figure}

\begin{figure}
\hspace{-0.1cm}\epsffile{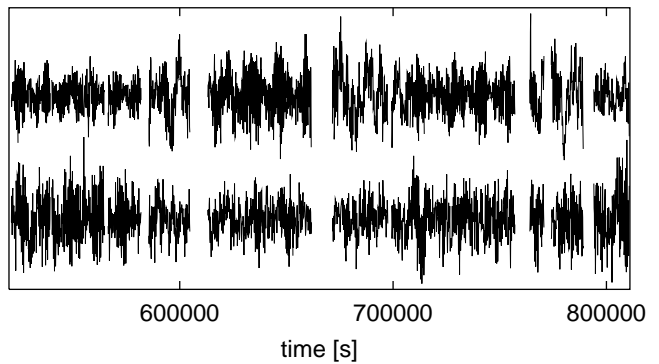}
\caption[]{\label{fig:esur}
  The down--sampled data set E with one corresponding surrogate. Gaps
  of different sizes prevents reasonable interpolation.  }
\end{figure}

Finally, the test is applied to experimental data. Data Set E of the
Santa Fe time series contest is a set of measurements of the
time--integrated intensity of light observed from a variable star. It
consists of 17 parts with different numbers of points, the time range
of which partly overlaps and partly shows gaps. Inside the blocks, the
data is evenly sampled with $\Delta=10$ s. Special interest in low
frequency components makes it desirable to consider the time series as
a whole. The Lomb periodogram of the data set is shown in
Fig.~\ref{fig:elomb}.

For the surrogate test we further down--sampled the data by
integrating over 12 successive measurements. Therefore, surrogates
could be generated in reasonable time. The resulting time series is
$N=2260$ points long with $\Delta=120$ s except for 9 gaps taking up
to 10000 s, as shown in Fig.~\ref{fig:esur}. The Lomb periodogram is
calculated at $N_f=1130$ frequencies with up to $\nu_{max}=1/240$ Hz.
The value for the time--reversibility statistic $\gamma = -0.56 \times
10^{-7}$ s$^{-3}$ of the considered time series does not lie outside
the interval $\gamma \in [-13 \times 10^{-7}$ s$^{-3},29 \times
10^{-7}$ s$^{-3}]$ spanned surrogates, and thus the null hypothesis
cannot be rejected. One surrogate time series is also shown in
Fig.~\ref{fig:esur}.

Spurious high frequency components could be introduced by
discrepancies in the overlapping parts of the recording. To deal with
that problem we deleted points from one of the two parts and repeated
the test. No significant differences in the surrogates and its values
for the test statistics were observed.

Taking a closer look at the individual parts shows considerable
differences in their autocorrelations, which makes it dangerous to
consider the whole data set as stationary. In contrast, the generated
surrogates are stationary by construction. If one could detect
significant differences with a nonlinear statistic, non--stationarity
would be an equally likely explanation as nonlinearity.

In this paper we presented a method for a test for nonlinearity for
time series with uneven time intervals. Such a test consists of two
main steps: generating surrogate data and calculating test statistics.
The new method is able to achieve the first step using the constrained
randomization scheme proposed in Ref.~\cite{anneal_surr}. We offered
only a first attempt on the second problem. More powerful test
statistics are likely to be derivable from current nonlinear time
series methods.

We like to thank Daniel Kaplan, James Theiler, Peter Grassberger and
Holger Kantz for useful discussions. This work was supported by the
SFB 237 of the Deutsche Forschungsgemeinschaft.

\end{document}